\documentclass{iaus}
\usepackage{graphicx,natbib}
\usepackage{bournaud}

\title[Star formation in galaxy mergers]
{Star formation in galaxy mergers:\\ ISM turbulence, dense gas excess, and scaling relations for disks and starbusts}

\author[Bournaud et al.]
{Fr\'ed\'eric Bournaud, Leila C. Powell, Damien Chapon \and Romain Teyssier}

\affiliation{CEA, IRFU, SAp -- CEA Saclay, F-91191 Gif-sur-Yvette, France.\\email: {\tt frederic.bournaud@cea.fr}}

\pubyear{2010}
\volume{271}  %% insert here IAU Symposium No.
\pagerange{}
\jname{Astrophysical Dynamics: From Stars to Galaxies}
\editors{N. Brummell, A. S. Brun, M. S. Miesch \& Y. Ponty}
\begin{document}

\maketitle

\begin{abstract}
Galaxy interactions and mergers play a significant, but still debated and poorly understood role in the star formation history of galaxies. Numerical and theoretical models cannot yet explain the main properties of merger-induced starbursts, including their intensity and their spatial extent. Usually, the mechanism invoked in merger-induced starbursts is a global inflow of gas towards the central kpc, resulting in a nuclear starburst. We show here, using high-resolution AMR simulations and comparing to observations of the gas component in mergers, that the triggering of starbursts also results from increased ISM turbulence and velocity dispersions in interacting systems. This forms cold gas that are denser and more massive than in quiescent disk galaxies. The fraction of dense cold gas largely increases, modifying the global density distribution of these systems, and efficient star formation results. Because the starbursting activity is not just from a global compacting of the gas to higher average surface densities, but also from higher turbulence and fragmentation into massive and dense clouds, merging systems can enter a different regime of star formation compared to quiescent disk galaxies. This is in quantitative agreement with recent observations suggesting that disk galaxies and starbursting systems are not the low-activity end and high-activity end of a single regime, but actually follow different scaling relations for their star formation. 
\keywords{galaxies: formation, galaxies: mergers, galaxies : star formation}
\end{abstract}

\firstsection 

\section{Introduction: Galaxy mergers as triggers of star formation}

Evidence that interactions and mergers can strongly trigger the star formation activity of galaxies has been observed for more than two decades \citep[e.g.,][]{sanders88}. Nevertheless, both the underlying mechanisms the role of mergers in the cosmic budget of star formation remain largely unknown.
Mergers have long appeared to potentially dominate the star formation history of the Universe. Starbursting galaxies with specific star formation rates much above the average, such as Ultraluminous Infrared Galaxies (ULIRGs), are almost exclusively major interactions and mergers -- in the nearby Universe (see \citealt{duc97}), not necessarily at high redshift. The decrease in the cosmic density of star formation between $z=1$ and $z=0$ \citep{lefloch05} may follow, and might even result from, the decrease in the galaxy merger rate. Other studies based on disturbed UV morphologies and/or kinematics also suggested high merger fractions among star forming galaxies at $z\sim1$ or even below \citep{hammer05}. 

However, disturbed morphologies and kinematics can also arise internally, without interactions, especially at high redshift when disk galaxies are wildly unstable, clumpy and irregular \citep{E07, FS09}. Merger-induced ULIRG-like starbursts may also be less important in the cosmic budget than the more moderate but more numerous objects with internally sustained star-formation, as is possibly the case for most high-redshift LIRGs \citep[e.g., ][]{daddi10a}. Recent studies aimed at accurately distinguishing the signatures of mergers from internal evolution actually suggest that major interactions and mergers account only for a small fraction of the cosmic star formation history \citep{jogee09, robaina09}. Moreover, observational estimates appear to be quite dependent on the chosen tracers of star formation. For instance, mergers are probably a more important trigger of dust-obscured star formation than of general star formation, and thus the fraction of mergers among objects with high infrared-traced star formation rate can be higher \citep[as discussed by][]{robaina09}.

Numerical simulations are then required to understand the mechanisms leading to starbursting activity in mergers, but also to further probe the contribution of mergers to cosmic star formation since observational estimates remain uncertain. Section~2 reviews the standard knowledge on merger-induced star formation, which is mostly based on ``sub-grid'' modeling: all steps from the formation of cold/dense gas clouds to the formation of actual stars remain unresolved, and described with arbitrary recipes. This standard understanding cannot account for some general observationed features that we briefly review. Section~3 presents a new generation of high-resolution models in which the first steps of galactic-scale star formation, namely ISM turbulence and cold/dense gas cloud formation, are explicitly resolved using high-resolution codes. Based on this, we provide a substantially different explanation for interaction-triggered star formation and show that it could better account for recent observations of disk galaxies and starbursting mergers.

\section{Standard models versus observations}
\subsection{Merger-induced gas flows}

In an interacting galaxy pair, the gas content of one galaxy undergoes tidal forces from the companion galaxy, but this is actually not the main direct driver of the gas response. The gas distribution becomes non-axisymmetric, because of the asymmetry in the gravitational field itself induced by the companion. This results in gravitational torquing of the gas. - a thorough description of the mechanism can be found in \citet[][Section~2 and Figure~1]{bournaud-proc}. Gas initially inside the corotation radius (typically a radius of a few kpc) undergoes negative gravity torques and flows inwards in a more and more concentrated central component. Gas outside the corotation, i.e. initially in the outer disk, undergoes positive gravity torques and gains angular momentum, forming long tidal tails.

\subsection{Nuclear starbursts}

Gravitational torquing in the inner disk increases the gas concentration in the central kiloparsec or so. Any model for star formation will then predict an increase of the star formation rate (global Schmidt-Kennicutt law, models based on cloud-cloud collisions, etc). The result is thus a centralized or nuclear starburst. As the driving process is gravitational torquing, early restricted three-body models could already describe the effect \citep{TT72}. Later models have added extra physical ingredients leading to more accurate predictions on the star formation activity: self-gravity \citep{BH92}, hydrodynamics and feedback processes \citep{mihos, cox-ratios}, etc.

A large library of SPH simulations of galaxy mergers, in which the driving process is mostly the one presented above (tidal torquing of gaseous disk) was performed and analyzed by \citep{dimatteo07,dimatteo08}. This study highlighted various statistical properties of merger-induced starbursts. In particular, it showed that some specific cases can lead to very strong starbursts with star formation rate (SFRs) are increased by factors of 10--100 or more, but that on average the enhancement of the SFR in a random galaxy collision is only a factor of a few (3--4 being the median factor) and only lasts 200-400~Myr.  These results were confirmed with code comparisons, and found to be independent of the adopted sub-grid model for star formation \citep{dimatteo08}. Models including an external tidal field to simulate the effect of a large group or cluster found that the merger-induced starbursts could be somewhat more efficient in such cases -- but the SFR increase remains in general below a factor of 10 \citep{MB08}.

\subsection{Theoretical predictions versus observations}

{\bf \noindent The intensity of merger-induced starbursts}
Numerical simulations reproducing the interaction-induced inflow of gas and resulting nuclear starbursts can sometimes trigger very strong starbursts, but in general the SFR enhancement peaks at 3--4 times the sum of the SFRs of the two pre-merger galaxies. This factor of 3--4 seems in good agreement with the most recent observational estimates \citep[e.g.]{jogee09, robaina09}. In fact, there is substantial disagreement: the factor 3--4 in simulation samples is the peak amplification of the SFR in equal-mass mergers. In observations, it is the average factor found at random (observed) instant of interactions, and in mergers that are "major" ones but not strictly equal-mass ones. Given that typical duration of a merger is at least twice longer than the starburst activity in the models, and that unequal-mass mergers make substantially weaker starbursts \citep{cox-ratios}, one would need a peak SFR enhancement factor of about 10--15 (as measured in simulations) to match the average enhancement of 3--4 found in observations. There is thus a substantial mismatch between the starbursting activity predicted by existing samples of galaxy mergers, and that observed in the real Universe -- although the observational estimates remain debated and may depend on redshift.

{\bf \noindent The spatial extent of merger-induced starbursts}

A disagreement between these 'standard' models/theories and observations of mergers is also found in the geometry and spatial extent of star formation. The very strongest merger-induced starbursts (ULIRGs) are centrally-concentrated, but there are many examples of significant merger-induced starbursts that are spatially extended. A well-known example are the Antennae galaxies, where the burst of star formation proceeds, for a large fraction, in a few big star-forming clumps in extended disks and in the bridge between the two galaxies \citep{wang04}. Another well-known example is the IC2163/NGC2207 pair, which has an extended starburst in big gas clouds with remarkably high gas velocity dispersions \citep{E95}. There are many other examples of extended star formation in merging systems \citep{cullen06,smith}. 

Quantitative comparisons of the extent of star formation in observations to that predicted by "standard" models have shown a significant disagreement \citet{barnes04,chienbarnes10}. These authors also suggested that a sub-grid model of shock-induced star formation may better account for the spatial extent of merger-induced star formation (see also Saitoh et al. 2009).

\medskip

The standard mechanism for merger-induced star formation, as reproduced in low-resolution simulations, certainly takes place in real mergers -- signs of nuclear merger-driven starbursts are a plenty. But it seems impossible to explain the typical intensity of merger-induced starbursts and their often relatively larger spatial extend, based on these "standard" models. Some relatively old observations may actually hold the key to correctly understanding merger-induced star formation:

\subsection{A key observation: ISM turbulence in interacting galaxies}

The cold gas component of interacting galaxies has high velocity dispersions, which can reach a few tens of km~s$^{-1}$. This has been noticed in the 90s in at least two well-resolved interacting and merging pairs \citep{irwin94,E95}. Many interacting pairs and mergers in \citet{green10} also have large velocity dispersions (in the ionized gas). More generally, massive star clusters in merging systems suggest the Jeans mass is high, which is indicative of high velocity dispersions. A typical number could be a factor 4 of increase for the cold gas velocity dispersion in equal-mass mergers. An interesting question will be understand whether these dispersions result from the tidal interaction and are a trigger star formation, or whether they just result from the starburst and associated feedback effects.

Traditional SPH simulations model a relatively warm gas for the ISM, because the limited spatial resolution translates into a minimal temperature under which gas cooling should not be modeled (it would generate artificial instabilities, \citealt{truelove}). Modeling gas cooling substantially below $10^4$~K requires "hydrodynamic resolutions"\footnote{i.e., the average smoothing length in SPH codes, or the smallest cell size in AMR codes} better than 100~pc. Cooling down to 100~K and below can be modeled only at resolutions of a few pc. The vast majority of existing merger simulations hence have a sound speed of at least 10~km/s and cannot explicitly treat the supersonic turbulence that characterizes most of the mass in the real ISM \citep{burkert-ism}. Turbulent speeds in nearby disk galaxies are of  5-10~km/s, for sound speed of the order of 1-2 km/s in molecular clouds, i.e. turbulent Mach numbers up to a few. These are even higher in high-redshift disks \citep{FS09}, and in mergers (references above), but not necessarily for the same reason.

Increased ISM turbulence in galaxy mergers is thus absent from the modeling used in most hydrodynamic simulations to date. Some particle-based models have nevertheless been successful in reproducing these increased gas dispersions \citep{E93,struck97,bournaud08}, indicating that it is a consequence of the tidal interaction which triggers non-circular motions, rather than a consequence of starbursts and feedback. It should then arise spontaneously in hydrodynamic models, if these a capable of modeling gas below $10^{3-4}$K.

\section{Merger-induced starbursts with resolved ISM turbulence and cluster star formation}

\subsection{High-resolution AMR simulations}
Adaptive Mesh Refinement (AMR) codes allow hydrodynamic calculations to be performed at very high resolution on adaptive-resolution grids. The resolution is not high everywhere, but the general philosophy is to keep the Jeans length permanently resolved until the smallest cell size is reached. That is, the critical process in the collapse of dense star-forming clouds, namely the Jeans instability (or Toomre instability in a rotating disk) is constantly resolved up to a typical scale given by the smallest cell size, or a small multiple of it. 
 
AMR simulations of whole galaxies have recently reached resolutions of a few pc for disk galaxies \citep{agertz,tasker}, and even 0.8~pc lastly \citep{bournaud-lmc}. Such techniques have been first employed to model ISM dynamics and star formation in galaxy mergers by \citet{kim,teyssier10}.
 
We here study the properties of star formation in a sample of a few AMR simulations of 1:1 mergers of Milky Way-type spiral galaxies, performed with a resolution of 4.5~pc and a barotropic cooling model down to $\sim$50~K, technically similar to the isolated disk simulation described in \citep{bournaud-lmc}. Star formation takes place above a fixed density threshold and is modeled with a local Schmidt law, i.e. the local star formation rate density in each grid cell is $\rho_{\mathrm SFR} = \epsilon_{\mathrm ff} \rho / t_{\mathrm ff} \propto \rho^{1.5}$ where $\rho$ is the local gas density and $t_ff$ the gravitational free-fall time, and the efficiency $\epsilon_{ff}$ is a fixed parameter. Supernova feedback is included. Further details and results for whole sample will be presented in Powell et al. (in preparation). An individual model of this type (but at lower resolution and without feedback), matching the morphology and star formation properties of the Antennae galaxies, was presented in \citealt{teyssier10}.
  \smallskip

\begin{SCfigure}[][t] 
\includegraphics[width=7cm]{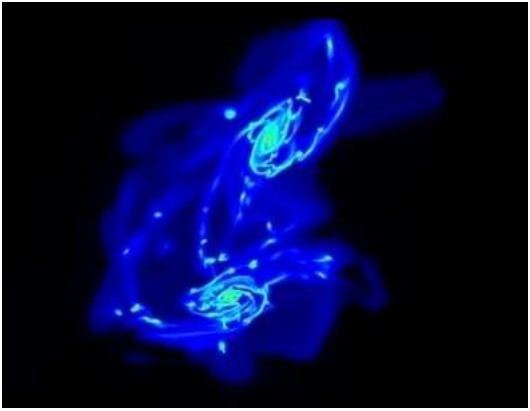}
\caption{\label{fs} Surface density of the cold gas component in an AMR merger simulation at 4.5~pc resolution. Dense gas clouds are more numerous and more massive than in similar models of isolated disk galaxies, owing to strong non-circular motions in interacting systems.}
\end{SCfigure} 
 
\subsection{Starburst properties}

In the models, the pre-merger isolated spiral galaxies spontaneously develop ISM turbulence at a about 10~km~s$^{-1}$ under the effect of gravitational instabilities (and/or feedback), and most star formation takes place in dense complexes of dense gas along spiral arms. In some sense, the large-scale star formation process is not entirely sub-grid anymore in these simulations, as the first steps of star formation, namely are the development of ISM turbulence and the formation of dense molecular gas clouds in this turbulent ISM, are explicitly captured -- the subsequent steps of star formation at smaller scales, inside the densest parts of these cold clouds, remain sub-grid.

A merger simulation is shown in Figure~\ref{fs}. The mass-weighted average of the gas velocity dispersion reaches $\sim 30$~km/s. This strong turbulence is consistent with the observations reviewed above. It induces numerous local shocks that increase the local gas density, which in turn triggers the collapse of gas into cold clouds. Also, gas clouds become more massive and denser than in the pre-merger spiral galaxies. The fraction of gas that is dense-enough and cold-enough to form star increases, and the timescale for star formation in these dense gas entities (the gravitational free-fall time) becomes shorter. As a result, the total SFR becomes several times higher than it was in the pre-interaction pair of galaxies. The standard process of merger-induced gas inflow towards the central kpc or so is also present, but the timescale is substantially longer, so this process dominates the triggering the star formation by enhancing the global gas density only in the late stages of the merger (Fig.~\ref{fs}).

\begin{SCfigure}[][t]
\centering
\includegraphics[width=6cm]{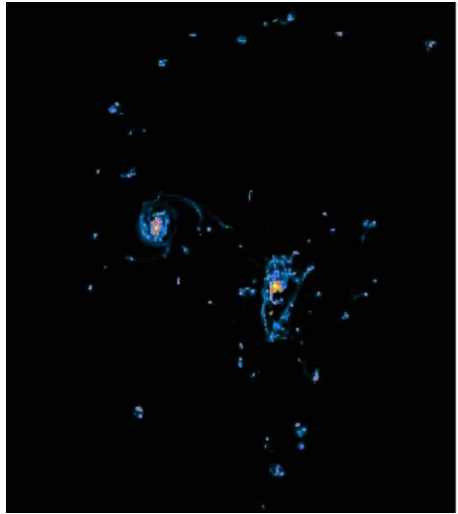}
\caption{\label{f1} Instantaneous star formation in an AMR merger model. Note the central nuclear starburst component from global gas inflows, but also the extended starburst component in massive clusters throughout the system. Dense knots of star formation along tidal tails are reminiscent of "beads on a string" star formation, as observed \citep[e.g.,][]{smith}.}
\end{SCfigure}

An example of star formation distribution is shown in Figure~\ref{f1}. Two consequences of modeling a cold turbulent ISM in merging galaxies are: (1) the peak intensity of the starburst can become stronger (as shown by Teyssier et al. 2010 for the Antennae) although this is not a systematical effect, and (2) the spatial extent of star formation in the starbursting phase is larger. The radius containing 50\% of the star formation rate (half-SFR radius) can more than double. This is because increased ISM turbulence is present throughout the disk, and triggers, through locally convergent flows and shocks, the collapse of efficiently star-forming clouds even at large distances from the nuclei. At least quantitatively, these results put the models in better agreement with observations. In these models, the increased ISM turbulence is also obtained without feedback, showing that it is not a consequence of the starbursting activity, but is rather driven by the tidal forces in the interaction.

\subsection{Dense gas fractions and star formation scaling relations}

{\bf \noindent Density PDFs and galactic-scaled star formation}

The connection between ISM turbulence, dense gas phases and star formation is best understood by examining the probability distribution function of the local gas density (density PDF). In its mass-weighted (resp. volume-weighted) version, the density PDF represents the mass (volume) fraction of the ISM in bins of density. We here use mass-weighted versions. 
 
In an system at equilibrium (e.g. isolated disk galaxy), supersonic turbulence generates a log-normal density PDF \citep{wada02}. Only small mass fractions of the whole ISM are at found number densities below $\sim 1$~cm$^{-3}$ or in very high density regions above, say, $10^4$~cm$^{-3}$, most mass being at ten to a few hundreds of atoms per cm$^{3}$ (typically observed as HI or CO-traced molecular gas). The width of the log-normal PDF depends on various factors, but at first order the main dependence is on the turbulent Mach number \citep{K07}. A more turbulent ISM will have a larger density spread in its density PDF, because the turbulent flows will sometimes diverge and create low-density holes, and sometimes converge (or even shock) into very dense structures that can further collapse gravitationally. Note that log-normal density PDFs are observed, at least at the scale of molecular complexes \citep{alves}. 
 
Our simulations of Milky Way-like disk galaxies do have quasi log-normal PDFs (see example on Figure~3). The PDFs are not exactly log-normal because the radial density gradient in the disk induces some deviations (the PDF is expected to be exactly log-normal only at fixed average density). Also, the spatial resolution limit converts into a density limit at which the density PDF is truncated, and which corresponds to the smallest/densest entities that can be resolved. The maximal density resolution in the present merger models is around $10^{6}$~cm$^{-3}$, but parsec-scale resolution can capture even higher densities (as in Bournaud et al. 2010).
 
 \medskip
 
The density PDF is a useful tool to describe the star-forming activity of a galaxy at a given gas content. For a simplified description, one can consider that star formation takes place only in the densest gas phases (i. e. above some density threshold), and that in these dense regions the local star formation rate follows, for instance, a fixed efficiency per free-fall time: $\rho_{\mathrm SFR} = \epsilon_{\mathrm ff} \rho_{\mathrm gas} / t_{\mathrm ff} \propto \rho_{\mathrm gas}^{1.5}$ (see detailed theory in Elmegreen 2002 and \citealt{K07}). Even if the local rate of star formation follows a different prescription than this purely density-dependent model (which is physically motivated by the gravitational collapse timescale), then the first step remains that star formation proceeds only in the densest gas phases. Thus, the fraction of dense gas remains a key parameter for the global star formation activity of galaxies.
 \smallskip
 
{\bf \noindent Dense gas excess and starbursts in mergers}
 
\begin{SCfigure}[][t]
\centering
\includegraphics[width=4.5cm]{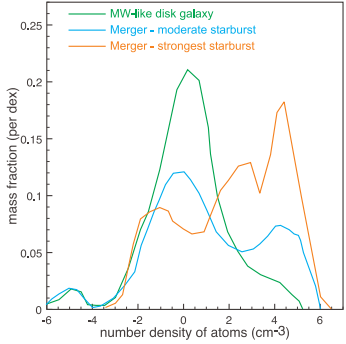}
\caption{\label{f2} Gas density PDFs (see text for details) of an isolated disk galaxy and two starbursting mergers, in our sample of high-resolution AMR simulations.}
\end{SCfigure}

Representative density PDFs are shown in our models on Figure~\ref{f2} for an isolated spiral galaxy, a moderately starbursting merger (SFR increased by a factor $\sim 3$ compared to the two pre-merger galaxies taken separately), and a stronger merger-induced starburst (factor $\sim 10$).

The PDFs of merging galaxies have a larger width than those isolated disk galaxies, as expected from the higher turbulent speeds that result from the tidal interaction. These PDFs are not necessarily log-normal\footnote{presumably because the turbulent motions do not follow a quasi-isotropic cascade and/or some of the involved non-circular motions are not really cascading turbulence}, but they clearly show a substantial excess of dense gas in starbursting mergers. Such density PDFs naturally imply high SFRs, idenpendently from the local SFR prescription, since the fraction of efficiently star-forming gas is high.

High fractions of dense gas in mergers were already proposed by \citet{juneau09}, based on detailed post-processing of merger simulations aimed at re-constructing dense molecular gas phases not resolved in these simulations. Here we obtain a qualitatively similar conclusion using simulations that explicitly resolve turbulent motions, local shocks and small-scale instabilities in cold ISM phases. This excess of dense gas should have signatures in molecular line ratios (as also observed by Juneau et al. 2009). If, in a rough approach, we assume that low-$J$ CO lines are excited for densities of 100~cm$^{-3}$ and above, and HCN lines for densities of $10^4$~cm$^{-3}$ and above, then the HCN/CO line ratios could be up to 5-10 times higher in the starbursting phases of major galaxy mergers. Simulations with an somewhat higher resolution would actually be desirable to accurately quantify the emission of dense molecular tracers.
 \smallskip

{\bf \noindent Star formation scaling relations}

The interpretation of merger-induced starbursts proposed from our high-resolution models is that it is not just a global gas inflow that increases the average gas density and increases the SFR, but also that there are strong non-circular motions, high turbulent velocity dispersions, causing many small-scale convergent flows and local shocks, that in turn initiate the collapse of dense star-forming clouds with high Jeans masses. The former ''standard'' process does take place, but the later can be equally important especially in the early phases of mergers.

We here note $\Sigma_{\mathrm gas}$ the average gas surface density of a galaxy. This is the quantity that observers would typically derive from the total gas mass and half-light radius, or similar quantities. The second mechanism above is a way to increase the SFR of a system, and its SFR surface density $\Sigma_{\mathrm SFR}$, without necessarily increasing its average $\Sigma_{\mathrm gas}$. Actually in our merger models $\Sigma_{\mathrm gas}$ does increase (as there are global merger-induced gas inflows), but $\Sigma_{\mathrm SFR}$ increases in larger proportions (as the starburst is not just from the global merger-induced inflow but also from the exacerbated fragmentation of high-dispersion gas). Going back to the density PDFs shown previously, one can note that the fraction of very dense gas (say, in the $\sim 10^{4-6}$~cm$^{-3}$ range) can increase by a factor of 10--20 in mergers while the average surface density $\Sigma_{\mathrm gas}$ increases by a factor 3--5 (see also on Figure~4). As a consequence, the $\Sigma_{\mathrm SFR}$ activity of these systems is unexpectedly high compared to their average surface density $\Sigma_{\mathrm gas}$.  
  
\begin{figure}
\centering
\includegraphics[width=5.5cm]{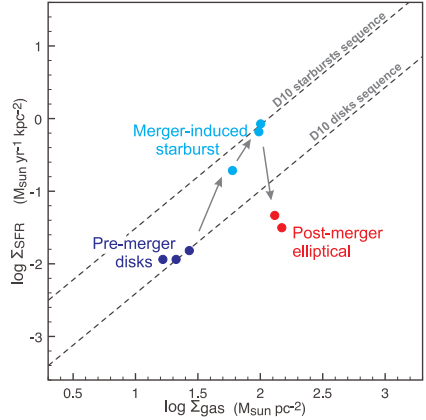}
\caption{Kennicutt-Schmidt diagram showing the evolution of a major merger simulation in the $ \left( \Sigma_{\mathrm gas} ; \Sigma_{\mathrm SFR} \right)$ plane (see text for details). The two dashed lines are the two star formation "laws" for disks and starbursts proposed by Daddi et al. (2010b). The pre-merger isolated disks evolve on the "disk sequence". The interaction and merger trigger a starburst not just through global gas inflows increasing the global averaged gas density, but also by increasing ISM turbulence and forming an excess of dense gas in massive cold clouds. The system then evolves towards the "starburst" mode as observed, while the increased fractions of dense gas could independently suggest a high excitation of molecular lines. The post-merger early-type galaxy settles back on a quiescent or even deficient mode. }\label{f3}
\end{figure}
 
Figure~4 shows the evolution of a system throughout a merger simulation in the  
$ \left( \Sigma_{\mathrm gas} ; \Sigma_{\mathrm SFR} \right)$ plane. While our pre-merger spiral galaxy models lie on the standard Kennicutt relation, starbursting mergers have high $\Sigma_{\mathrm SFR} / \Sigma_{\mathrm gas}$ ratios. This is in agreement with observational suggestions that quiescent disks and starbursting mergers do not follow the same scaling relations for star formation, but could actually display two different star formation "laws" \citep{daddi10b, genzel10}. The offset between the disk and merger sequences proposed by Daddi et al. (2010b) is quantitatively recovered in our simulations (Figure~4). Post-starburst, post-merger systems lie back on the quiescent sequence, or even somewhat below it: these systems contain some dense gas which is somewhat stabilized by the stellar spheroid. This is another example of a "morphological quenching" effect in early-type galaxies \citep{martig09}, and the location of our post-merger early-type galaxies in the $ \left( \Sigma_{\mathrm gas} ; \Sigma_{\mathrm SFR} \right)$ diagram may be consistent with observations of nearby ellipticals (\citealt{crocker10}, but see \citealt{fabello}). 

The proposal that disks and mergers follow two different regimes of star formation by Daddi et al. (2010b) and Genzel et al. (2010) relies for a part (but not entirely) on the assumption that different CO luminosity-to-molecular gas mass conversion factors apply in quiescent disks and starbursting mergers. Interestingly, our simulations recover the two regimes of star formation without any assumption on such conversion factors since gas masses are directly known. But at the same time, excess of dense gas found in these merger models suggests that the excitation of CO lines would naturally be higher in mergers/starbursting phases (although this needs to be quantified in the models), which would mean that the assumption of different conversion factors by Daddi et al. and Genzel et al. could be physically justified. High molecular gas excitation in SubMillimeter Galaxies (SMGs, Tacconi et al. 1998) could then naturally result if these are starbursting major mergers with high gas surface densities and a clumpy turbulent ISM \citep[e.g.][]{narayanan, bournaud10}.

\section{Summary}

The results presented here were based mostly on low-redshift merger simulations. A recent set of high-redshift merger simulations with AMR is presented in Bournaud et al. (2011), and shows similar increase in the ISM velocity dispersions and clumpiness in mergers, with extended starbursts and high $\Sigma_{\mathrm SFR}/\Sigma_{\mathrm gas}$ ratios.

Galaxy interactions and mergers play a significant but still debated and poorly understood role in the star formation history of galaxies. Numerical and theoretical models have significant difficulties in accounting for the properties of merger-induced starbursts, including their intensity and their spatial extent. Usually, the mechanism invoked to explain the triggering of star formation by mergers is a global inflow of gas towards the central kpc, resulting in a nuclear starburst. We show here, using high-resolution AMR simulations and comparing to observations of the gas component in mergers, that the triggering of starbursts also results from increased ISM turbulence and velocity dispersions in interacting systems. This results in the formation and collapse of dense and massive gas clouds in the regions of convergent flows and local shocks, these clouds being denser, more massive and/or more numerous than in quiescent disk galaxies. The fraction of dense cold gas largely increases, modifying the global density distribution of these systems, and efficient star formation results. Because the starbursting activity is not just from a global compacting of the gas to high average surface densities, but also from higher turbulence and fragmentation into massive and dense clouds, merging systems can enter a different regime of star formation compared to quiescent disk galaxies. This is in quantitative agreement with recent observations suggesting that disk galaxies and starbursting systems are not the low activity end and high activity end of a single regime, but actually follow different scaling relations for their star formation.

\acknowledgements
FB acknowledges numerous discussions on gas dynamics and star formation in galaxy mergers with Bruce Elmegreen, Pierre-Alain Duc, Emanuele Daddi and many others, and is grateful to the organizers of the symposium for the exciting discussions along a wide range of astrophysical problems.

\end{document}